\documentclass[11pt,letterpaper]{article}
\pdfoutput=1
\usepackage{mathtools}
\usepackage{axodraw2}
\usepackage{amscd}
\usepackage{amsmath}
\usepackage{slashed}
\usepackage{gensymb}
\usepackage{jheppub} 

\title{Third Family Hypercharge Model for $R_{K^{(\ast)}}$ and Aspects of the
 Fermion Mass Problem}   

\author{B.C. Allanach,}
\author{Joe Davighi\footnote{Corresponding author.}}
\affiliation{DAMTP, University of Cambridge, Wilberforce Road, Cambridge, 
CB3 0WA, United Kingdom}

\emailAdd{B.C.Allanach@damtp.cam.ac.uk}
\emailAdd{jed60@cam.ac.uk}
\preprint{DAMTP-2018-30}
\abstract{We present a model  to explain LHCb's recent measurements of $R_K$
  and $R_{K^{\ast}}$ based on an anomaly-free, spontaneously-broken $U(1)_F$ gauge symmetry, without any fermionic
fields beyond those of the Standard Model (SM). The model explains the
hierarchical heaviness of the third family and the smallness of quark
mixing. The $U(1)_F$
charges of the third family of SM fields and the Higgs doublet are
set equal to
their respective hypercharges.
A heavy $Z^\prime$ particle with flavour-dependent couplings can modify the
 $[\overline{b_L} \gamma^\rho 
 s_L][\overline{\mu_L} \gamma_\rho 
\mu_L]$ effective vertex in the 
desired 
way. The $Z^\prime$ contribution to
$B_s-\overline{B_s}$ mixing is suppressed by a small mixing angle connected to
 $V_{ts}$, 
making the constraint coming from its measurement easier to satisfy.
The model can explain $R_K$ and $R_{K^{\ast}}$ whilst simultaneously
passing other constraints, including measurements of the lepton flavour
universality of $Z$ couplings. 
}

\begin{document} 
\maketitle
\flushbottom

\section{Introduction}
Recent measurements of semi-leptonic $B-$meson decays by the LHCb
collaboration in LHC Run I~\cite{Aaij:2014ora,Aaij:2017vbb} suggest that they
violate electron-muon 
universality more 
than is predicted by the SM~\cite{Hiller:2003js}.
The primary evidence comes from $B\rightarrow K^{(\ast)}l^+l^-$ where $l \in \{
e,\mu \}$, as encapsulated by the $R_K$ and $R_{K^\ast}$ parameters:
\begin{equation}
R_K = \frac{BR(B\rightarrow K \mu^+\mu^-)}{BR(B\rightarrow
  K e^+e^-)}, \qquad
R_{K^\ast} = \frac{BR(B\rightarrow K^{\ast}\mu^+\mu^-)}{BR(B\rightarrow
  K^{\ast}e^+e^-)}.  \label{Rs}
\end{equation}
There are three such discrepant measurements: $R_K$ in a di-lepton invariant mass
squared bin of $Q^2 \in [1,\ 6]$ GeV$^2$ which disagrees with the SM at
2.6$\sigma$, $R_{K^\ast}( Q^2 \in [0.045,\ 1.1]\text{~GeV}^2)$ which has a
2.2$\sigma$ level discrepancy with the SM prediction and $R_{K^\ast}(Q^2 \in
[1.1,\ 6]\text{~GeV}^2)$, which is at odds with the SM prediction at the
2.5$\sigma$ level. Each individual measurement is not especially significant,
but their combination is around the 4$\sigma$ level. 
The prediction of $R_K$ and $R_{K^\ast}$ in the SM is particularly clean,
since the theoretical uncertainties cancel nicely in the ratios to leave only
a very small overall uncertainty. 
Analyses of Run II data are eagerly awaited, as are similar measurements from
BELLE II, but in the meantime many have tried to explain the discrepancies 
with a new physics effect. 
If we
quantify the effect of new physics by a change to the Wilson coefficient of a
single effective field theory (EFT) operator, 
\begin{equation}
\mathcal{O} \propto [\bar{s} \gamma_{\mu} P_L b] [\bar{\mu} \gamma^{\mu} P_X
\mu] \label{effOp}
\end{equation}
is found to fit $R_K$, $R_{K^\ast}$ and other $B-$physics
data~\cite{DAmico:2017mtc}, where $P_X=1$ (i.e.\ a vector-like coupling to
muon pairs) or  $P_X=P_L$ (a left-handed coupling to
muon pairs). $P_X=P_R$ does {\em not}\/
  provide a good fit.
This EFT operator may arise from integrating out some heavy new particle
which preferentially couples to muons rather than electrons. 
The fits indicate that the mass of the new particle is 31 TeV divided by the
square root of a product of two couplings. 
Since the couplings are flavour dependent, this raises the exciting
possibility of experimentally probing new physics {\em which could potentially
  explain the pattern of hierarchies in fermion masses and mixings}. 
In the
spirit of simplified model building, one begins by looking for a single new
particle to explain the data.
It is found at tree-level that this new
particle~\cite{Gripaios:2015gra}\footnote{Other approaches based on more complete model set-ups
 have been discussed, for example composite
 Higgs~\cite{Carmona:2015ena,Carmona:2017fsn}, composite
 leptoquark~\cite{Gripaios:2014tna}, or warped extra
 dimensional~\cite{GarciaGarcia:2016nvr,Carena:2018cow} 
 models.} could either be a flavour-dependent 
leptoquark or a
$Z^\prime$ with flavour dependent
couplings~\cite{Gauld:2013qba,Buras:2013dea,Buras:2013qja,Altmannshofer:2014cfa,Buras:2014yna,Crivellin:2015mga,Crivellin:2015lwa,Sierra:2015fma,Crivellin:2015era,Celis:2015ara,Greljo:2015mma,Altmannshofer:2015mqa,Allanach:2015gkd,Falkowski:2015zwa,Chiang:2016qov,Becirevic:2016zri,Boucenna:2016wpr,Boucenna:2016qad,Ko:2017lzd,Alonso:2017bff,Alonso:2017uky,1674-1137-42-3-033104,CHEN2018420,Faisel:2017glo,PhysRevD.97.115003,Bian:2017xzg,PhysRevD.97.075035,Duan:2018akc}.  
It is the latter possibility that we focus on in the present paper. 

Here, our {\em modus operandi}\/ is to incrementally model-build the $Z^\prime$
simplified model 
toward a more fundamental theory. One obvious choice is to take the $Z^\prime$
to be 
the heavy gauge boson from an underlying $U(1)_F$ flavoured gauge symmetry.\footnote{By the word ``flavoured'', we here mean that the gauge field for the $U(1)_F$ symmetry has flavour-dependent couplings to the SM fermions.} For example, in the Third Family Hypercharge Model, we will arrange the $U(1)_F$ charges of the SM fermions such that only the third
family is allowed a gauge-invariant Yukawa coupling at the renormalisable
level.\footnote{Third family hypercharge times another $U(1)$ gauge symmetry
(first two family hypercharge) were simultaneously broken to the diagonal
subgroup $U(1)_Y$ 
in Ref.~\cite{Chiang:2007sf}. The model was not connected to any $B-$anomalies 
(it was invented before the relevant measurements)
but it does address
some aspects 
of the fermion mass problem after adding either additional Higgs doublets or
an additional vector-like family to the SM.} 
The first two families and
neutrinos may
then acquire Yukawa couplings (or masses) at the
non-renormalisable level. 
Fermion mixing may also be generated by such
non-renormalisable terms, for example by the Froggatt-Nielsen mechanism~\cite{Froggatt:1978nt}. 
To be guaranteed a consistent quantum field theory, one 
chooses the charges such that no anomalies (including mixed anomalies and 
gauge-gravity anomalies) arise. This is
a highly non-trivial constraint on possible $U(1)_F$ charges. 

Our path then is clear: we wish to find anomaly-free combinations of $U(1)_F$
charges which predict that the terms in Eq.~\ref{effOp} are present, as well as
the third family Yukawa couplings (but no other Yukawa couplings at the
renormalisable level). 
We find that there is indeed such an anomaly-free set of $U(1)_F$ charges
which satisfies these criteria, without the need to introduce any additional
fermion fields beyond the SM content. Moreover, this solution to
the anomaly constraints is unique up to normalisation, with the $U(1)_F$
charge of each third family field proportional to its
hypercharge.\footnote{The space of solutions to the anomaly constraints has
    been recently studied in Ref. \cite{Ellis:2017nrp}, subject to the
    constraint that the right-handed down-type quarks all have vanishing
    $U(1)_F$ charge. With such a constraint, it is shown in
    Ref.~\cite{Ellis:2017nrp} that one must introduce additional fermions (which
    may be identified as dark matter candidates) to satisfy the anomaly
    equations. In the present work, we evade this conclusion by allowing $b_R$
    to have a $U(1)_F$ charge, and so we are able to find an anomaly-free set
    of charges with only SM fields.}
After this, we wish to make sure that the model passes all existing
constraints upon it, since it will necessarily predict additional $Z^\prime$
couplings to other fermions than just those in Eq.~\ref{effOp}: for example,
it should also be possible in the model to somehow reduce the
$Z^\prime$ coupling to $\bar s \gamma_\mu P_R b$, which fits the $B-$physics
data badly if it dominates over $\bar s \gamma_\mu P_L b$. 

Somewhat similar approaches (which also
  aim to connect the recent $B$-physics data with the hierarchies in the
  fermion masses or their mixing angles) have  been made in the literature. As
  we shall now discuss, our approach 
  is significantly different to these, both in its aims and in its
  construction.
In Refs.~\cite{Bonilla:2017lsq,Bhatia:2017tgo} 
    different anomaly-free sets of gauged
    $U(1)_F$ charges were found in models with additional fermionic SM singlet
    fields and an additional Higgs doublet. 
These models allow enough Yukawa couplings at the renormalisable level to
achieve quark 
and lepton mixing, as well as yielding the effective field theory operator in
Eq.~\ref{effOp}. 
Refs.~\cite{Alonso:2017uky,Alonso:2017bff} are similar in spirit to
Refs.~\cite{Bonilla:2017lsq,Bhatia:2017tgo}, except that mixing between the
third family and the first two is banned at the renormalisable level in
the former two papers. 
However, the models shed no light on the origin of
the hierarchy in fermion masses. 
Thus,
Refs.~\cite{Bonilla:2017lsq,Bhatia:2017tgo,Alonso:2017uky,Alonso:2017bff} are
quite different to our approach in which,
by requiring that only the third family fermions are allowed Yukawa
  couplings at the renormalisable level, we provide a possible explanation for
  the hierarchical heaviness of the third family and of the small size of
  quark mixing. 

In Ref.~\cite{Falkowski:2015zwa}, a spontaneously broken $U(2)_F \cong SU(2)_F \times U(1)_F$ flavour symmetry, in which the $U(1)_F$ subgroup is gauged, was used to
explain $R_{K^{(\ast)}}$ via the flavour-dependent interactions of the corresponding $Z^\prime$. 
By introducing scalar spurions that parametrize the $U(2)_F$ breaking, the authors of Ref.~\cite{Falkowski:2015zwa} are able to arrange appropriate power-law hierarchies for the fermion masses and mixing angles, from $\mathcal{O}(1)$
renormalisable fundamental couplings \'{a} la Froggatt-Nielsen ~\cite{Froggatt:1978nt}. 
The model fits the $R_{K^{(\ast)}}$ measurements by increasing the denominators in Eq.~\ref{Rs} whilst simultaneously
decreasing the numerators. Increasing the denominators is (by now) somewhat
disfavoured by global fits to various $B$ data. 
Furthermore, as is typical in Froggatt-Nielsen inspired model-building, the light generation quarks carry the largest charges under the flavoured $U(1)_F$ symmetry. Consequently, the $Z'$ boson in such a model couples most strongly to the valence quarks $u$, $d$, and $s$, and is therefore subject to stronger constraints from current data (for example from the high-$p_T$ dilepton tails ~\cite{Greljo:2017vvb} in $pp$ collisions).
Nonetheless, the model of Ref.~\cite{Falkowski:2015zwa} remains
similar in its aims to ours and, indeed, goes further into detail on the
fermion mass model-building by explicitly writing down the higher dimension operators responsible for the light fermion masses (and mixings), at the level of an effective description involving SM fields and spurions. 

The model proposed in Ref.~\cite{King:2018fcg}  also seeks to connect  the
$R_{K^{(*)}}$ measurements with the fermion mass hierarchies through a gauged
and spontaneously broken $U(1)_F$ symmetry with flavour-dependent
couplings. In that model, as well as an additional Higgs doublet, 
a vector-like fourth family of SM fermions was
introduced to produce the required operator Eq.~\ref{effOp}. The vector-like
fourth family is the only one charged under $U(1)_F$, meaning that
gauge anomalies are cancelled. 
Whilst the existence of a gauged $U(1)_F$ symmetry and a 
connection between fermion mass predictions and the $B-$discrepancies is in 
common 
with Refs.~\cite{King:2018fcg}, our model differs in that it is
anomaly-free without adding any matter fields or Higgs doublet fields to the
SM field content. In that sense, our model is a more minimal extension to the SM. 
An attempt has also been made to
  connect the $R_{K^{(*)}}$ measurements with the fermion mass hierarchies
  through a leptoquark model, rather than a $Z^{\prime}$, in
  Ref. \cite{Deppisch:2016qqd}.\footnote{In this leptoquark model, a {\em global}\/ ({\em
    i.e.}\/ not gauged) $U(1)_F$ symmetry is invoked. The assignment of $U(1)_F$
  charges to the SM fermions is arranged so as to produce mass hierarchies via
  the Froggatt-Nielsen mechanism, while the assignment of $U(1)_F$ charges to
  the leptoquarks leads to a hierarchy in the leptoquark couplings which is
  capable of explaining the $B$-physics data. Note that, in this model, the
  $U(1)_F$ charges of the leptoquarks can be chosen independently to the SM
  fermion charges, such that the explanations for the $B$-physics data and for
  the fermion masses effectively decouple into two independent
  explanations. In contrast, in our model, we only have the SM
  fermions, whose assignment of $U(1)_F$ charges can provide a {\em shared}\/
  explanation for both the $B$-physics data and the fermion masses, which is
  moreover anomaly-free.}

We note that another paper 
introduced a simplified\footnote{There have been ultra-simplified
  $Z^\prime$ models 
  in the literature, e.g.~\cite{Gauld:2013qba,Allanach:2015gkd}, where only
  couplings to muon flavoured leptons and 
  $\bar s b+H.c.$ have been considered. These do not preserve $SU(2)_L$.}
$Z^\prime$ model where the $Z^\prime$ coupled dominantly to left-handed bottom
quarks and to left-handed muons~\cite{Allanach:2017bta}. No attempt was made
to solve any anomaly
constraints or to explain aspects of the observed fermion masses and mixings,
and so we construct a more complete model here.

\section{Third Family Hypercharge Model}
In order to ban all Yukawa couplings except those of the third family,
we set the $U(1)_F$ charges of the first two families to zero but give the
Higgs $H$ a non-zero charge. 
With this constraint, the only set of charges that satisfies all of the
anomaly equations is the one where fermion charges of the third family are
proportional to their hypercharges. 
Since it is well known that hypercharges fit into grand unified groups such as
$SU(5)$ and $SO(10)$, 
the $SU(3) \times SU(2)_L \times U(1)_Y
  \times U(1)_F$ gauge symmetry
may be embedded 
within some larger non-abelian unified symmetry. We list the charges in
Table~\ref{tab:charges}. 
\begin{table}
\begin{center}
\begin{tabular}{|cccccc|}\hline
$F_{Q_i'}=0$ &  $F_{{u_R}_i'}=0$ & $F_{{d_R}_i'}=0$ & $F_{L_i'}=0$ & $F_{{e_R}_i'}=0$ &
 $F_H=-1/2$ \\
$F_{Q_3'}=1/6$ &  $F_{{u_R'}_3}=2/3$ & $F_{{d_R'}_3}=-1/3$ & $F_{L_3'}=-1/2$ &
                                                                               $F_{{e_R'}_3}=-1$ & 
  \\ \hline
\end{tabular}
\caption{\label{tab:charges} $U(1)_F$ charges of the fields in the Third
  Family Hypercharge Model, where $i \in \{1, 2\}$. All gauge
anomalies, mixed gauge anomalies and mixed gauge-gravity anomalies cancel. }
\end{center}
\end{table}
At the renormalisable level, the only allowed Yukawa couplings are
\begin{equation}
{\mathcal L}=Y_t  {\overline{{Q_3}_L'}} H t_R' + Y_b \overline{{Q_3'}_L} H^c b_R' + Y_\tau {\overline{{L_3}_L'}} H^c
\tau_R'  + H.c., \label{3famYuk}
\end{equation}
thus explaining the relative hierarchical heaviness of the third
family\footnote{$Y_t$, $Y_b$ and $Y_\tau$ are complex dimensionless Yukawa
  couplings.} once the neutral component of the Higgs doublet acquires its
vacuum expectation value $v$. Under the SM gauge symmetry $SU(3)\times SU(2)_L
\times 
U(1)_Y$, the fields transform as $H \sim(1, 2, -1/2)$,
$${Q_i}_L'\sim(3, 2, 1/6),\ 
{L_i}_L'\sim(1, 2, -1/2),\ 
{u_i}_R'\sim(3, 1, 2/3),\ 
{d_i}_R'\sim(3, 1, -1/3),\ 
{e_i}_R'\sim(1, 1, -1),$$
where we suppress gauge indices but not the family index $i \in \{1,2,3\}$
and $H^c=(H^+,\ -{H^0}^\ast)^T$.
Weak eigenbasis fermion fields are written with a prime, whereas fermion
fields in the mass eigenbasis shall be written without a prime. 
The only Yukawa terms allowed are precisely those in Eq.~\ref{3famYuk}. 
More detailed model building may provide estimates for neutrino and lighter
family
masses, and fermion mixings, which may come from non-renormalisable
operators. 
A small perturbation of Eq.~\ref{3famYuk} from such non-renormalisable
operators will necessarily predict small quark mixing.  
For now, we shall simply constrain fermion mixings and the masses of the first
two generations to be at their 
central measured values.

\subsection{Masses of gauge bosons and $Z-Z^\prime$ mixing} \label{ZZpmixing}

The $U(1)_F$ symmetry is assumed to be spontaneously broken by a SM singlet complex scalar
flavon, $\theta$. Its charge under $U(1)_F$ is $F_\theta\neq 0$, and we denote its
vacuum expectation value (VEV)
by $v_F$. We denote the original $U(1)_F$ gauge boson by $X$, reserving the name $Z^{\prime}$ for the physical boson (which is a mass eigenstate). The original $Z$ boson of the SM mixes with this $X$ boson 
to a small degree because the neutral component of $H$, which achieves a VEV $v$,
has both $U(1)_F$ and $SU(2)\times U(1)_Y$ quantum numbers.\footnote{We assume
  that the kinetic term for the gauge fields themselves, which should {\em a
    priori}\/ include an off-diagonal term mixing the two $U(1)$ gauge fields,
  has already been diagonalised.} 
Following Refs.~\cite{Bandyopadhyay:2018cwu,Duan:2018akc} (which examined some $Z-Z^\prime$
mixing constraints in different $SM\times U(1)$ models), the relevant mass
terms come from the kinetic terms of the scalar fields $H$ and $\theta$: 
\begin{equation}
\mathcal{L}_{H\theta K} = (D^\mu H)^{\dagger}(D_\mu H) + (D^\mu \theta)^{*}(D_\mu \theta), \label{scalarKineticTerms}
\end{equation}
where the covariant derivatives are
\begin{equation}
D_{\mu}H=\partial_\mu H+i\frac{g}{2}\left(\tau^a W^a_{\mu}-\frac{g'}{g}B_\mu - \frac{g_F}{g}X_\mu\right)H, \qquad D_{\mu}\theta=(\partial_\mu +iF_{\theta} g_F X_\mu)\theta,
\end{equation}
where, as usual, $g$ and $g'$ denote the gauge couplings for $SU(2)_L$ and $U(1)_Y$ respectively, and $g_F$ denotes the gauge coupling for $U(1)_F$.

Expanding the scalar fields about their VEVs in Eq.~\ref{scalarKineticTerms},
{\em viz.}\/ $H=(v+h(x),\ 0)^T/\sqrt{2}$ and
$\theta=(v_F+s(x))/\sqrt{2}$, leads to mass terms for the neutral gauge bosons
of the form $\mathcal{L}_{N,\text{mass}} = \frac{1}{2}{\bf A'_\mu}^T
\mathcal{M}^2_N {\bf A'_\mu}$, where ${\bf
  A'_\mu}=(B_\mu,W_\mu^3,X_\mu)^T$,\footnote{Here, the prime on ${\bf A'}_\mu$
  denotes that the gauge fields are in the $SU(3) \times SU(2)_L \times U(1)_Y
  \times U(1)_F$ eigenbasis.} and the mass matrix is
\begin{equation}
\mathcal{M}^2_N=\frac{v^2}{4}\left( \begin{array}{ccc}
g'^2 & -gg' & g'g_F \\
-gg' & g^2 & -gg_F \\
g'g_F & -gg_F & g_F^2(1+4F_\theta^2 r^2) \\
\end{array}\right),
\end{equation}
where $r\equiv v_F/v\gg 1$ is the ratio of the VEVs. One can check that the determinant of $\mathcal{M}^2_N$ vanishes, hence there remains a massless photon. Writing $\mathcal{L}_{N,\text{mass}} =- \frac{1}{2}{\bf A'_\mu}^T OO^T \mathcal{M}^2_N OO^T {\bf A'_\mu}$, where $O$ is an orthogonal matrix such that $O^T \mathcal{M}^2_N O=\text{diag}(0,M_Z^2,M_{Z^\prime}^2)$, we define the mass basis of physical neutral gauge bosons via $(A_\mu,Z_\mu,Z^{\prime}_\mu)^T\equiv{\bf A_\mu}=O^T {\bf A'_\mu}$. The orthogonal matrix $O$ can be written 
\begin{equation}
O=
\left( \begin{array}{ccc}
\cos\theta_w & -\sin\theta_w \cos\alpha_z & \sin\theta_w \sin\alpha_z \\
\sin\theta_w & \cos\theta_w \cos\alpha_z & -\cos\theta_w \sin\alpha_z \\
0 & \sin\alpha_z & \cos\alpha_z \\
\end{array}\right), \label{orthogonal}
\end{equation}
where $\theta_w$ is the Weinberg angle (such that $\tan\theta_w=g'/g$), and the $Z-Z^\prime$ mixing angle $\alpha_z$ is related to the masses of the $Z$ and $Z^{\prime}$ via the equation
$$(M_{Z^\prime}^2-M_Z^2)\sin 2\alpha_z = \frac{g g_F v^2}{2 \cos\theta_w}.$$
We shall now assume that the $Z^\prime$ is much heavier than the $Z$ boson, such that the mixing between them is small. In the (consistent) limit that $M_Z/M_Z^\prime\ll 1$ and $\sin\alpha_z \ll 1$, the masses of the heavy neutral gauge bosons are given by
\begin{equation}
M_Z\approx \frac{M_W}{\cos\theta_w}=M_W\frac{\sqrt{g^2+g'^2}}{g}, \qquad M_{Z^\prime}\approx M_W \frac{g_F\sqrt{1+4F_\theta^2 r^2}}{g},
\end{equation}
where $M_W =gv/2 $, and the mixing angle is
\begin{equation}
\sin\alpha_z \approx \frac{g_F}{\sqrt{g^2+g'^2}}\left(\frac{M_Z}{M_Z^\prime}\right)^2. \label{mixing}
\end{equation}
Recall that the ratio of VEVs $r=v_F/v$ is much larger than one, such that the $Z^\prime$ is indeed expected to be much heavier than the electroweak gauge bosons of the SM.

From the relation ${\bf A_\mu}=O^T {\bf A'_\mu}$, and Eq.~\ref{orthogonal}, one deduces that the photon remains the same linear combination of $B$ and $W^3$ as in the SM\@. The physical $Z$ boson, however, now contains a small admixture of the $X$ field:
\begin{equation}
Z_\mu = \cos\alpha_z \left( -\sin\theta_w B_\mu + \cos\theta_w
  W_\mu^3\right)+\sin\alpha_z X_{\mu}, \label{zBoson} 
\end{equation}
and so will inherit small flavour-changing corrections to its fermionic
couplings.
Thus, we must take the $Z$ boson mediated contributions into account when
calculating flavour violating effective operators. 
The $Z-Z^\prime$ mixing must be consistent with a constraint from LEP, as we shall see in
 \S~\ref{sec:constraints}. 

\subsection{$Z^\prime$ couplings to fermions \label{zPcouplings}}

We begin with the couplings of the $U(1)_F$ gauge boson $X_\mu$ to fermions in
the Lagrangian in the weak eigenbasis
\begin{equation}
{\mathcal L}_{X \psi} = -
g_F \left( 
\frac{1}{6}{\overline{{Q_3'}_L}} \gamma^\rho {Q_3'}_L -\frac{1}{2}
{\overline{{L_3'}_L}} \gamma^\rho {L_3'}_L -
{\overline{{e_3'}_R}} \gamma^\rho {e_3'}_R +\frac{2}{3}
{\overline{{u_3'}_R}} \gamma^\rho {u_3'}_R -\frac{1}{3}
{\overline{{d_3'}_R}} \gamma^\rho {d_3'}_R
\right) X_\rho, \label{Zpcouplings}
\end{equation}
where $g_F$ is the $U(1)_F$ gauge coupling. We  saw in \S~\ref{ZZpmixing} that
the $U(1)_F$ gauge boson $X$ is equal to the physical heavy gauge boson
$Z^\prime$ (which is a mass eigenstate) up to a small correction.  
In order to calculate the effects on the mass eigenbasis fields, we must
provide the connection to the weak eigenbasis for the fermions: the details
 and conventions are 
set out in Appendix~\ref{sec:weakToMass}. 
In the mass basis, using
Eqs.~\ref{orthogonal},\ref{mixing}, and~\ref{fermion rotations}, Eq.~\ref{Zpcouplings} becomes
\begin{eqnarray}
\mathcal{L}_{X \psi} =-&g_F&\left( 
\frac{1}{6}\overline{\bf u_L} \Lambda^{(u_L)} \gamma^\rho {\bf u_L} + 
\frac{1}{6}\overline{\bf d_L} \Lambda^{(d_L)} \gamma^\rho {\bf d_L}-\frac{1}{2}
\overline{\bf n_L} \Lambda^{(n_L)} \gamma^\rho {\bf n_L} -\frac{1}{2}
\overline{\bf e_L} \Lambda^{(e_L)} \gamma^\rho {\bf e_L}\right. \nonumber \\
&&\left. +\frac{2}{3}
\overline{\bf u_R} \Lambda^{(u_R)} \gamma^\rho {\bf u_R}-\frac{1}{3}
\overline{\bf d_R} \Lambda^{(d_R)} \gamma^\rho {\bf d_R}-
\overline{\bf e_R} \Lambda^{(e_R)} \gamma^\rho {\bf e_R}
\right)Z^\prime_{\rho},  
\label{secSU2}
  \end{eqnarray}
where each of the couplings is missing small $\mathcal{O}
\left({M_Z^2}/{{M_Z^\prime}^2}\right)$ terms induced by $Z-Z^\prime$
mixing, and
we have 
defined the 3 by 3 dimensionless Hermitian coupling matrices 
\begin{equation}
\Lambda^{(I)} \equiv V_{I}^\dagger \xi V_{I} ,
\label{lambdas}
\end{equation}
where
$I \in \{ u_L, d_L, e_L, \nu_L, u_R, d_R, e_R \}$ and
\begin{equation}
\xi = \left(\begin{array}{ccc}
0 & 0 & 0 \\ 0 & 0 & 0 \\ 0 & 0 & 1 \\
\end{array}\right).
\end{equation}
This completes our definition of the Third Family Hypercharge Model. 
Provided that $(V_{e_L})_{23}\neq 0$ and $(V_{d_L})_{23} \neq 0$,
Eq.~\ref{secSU2} contains couplings to $\overline{b_L} s_L+H.c.$ and $\overline{\mu_L}
\mu_L$, and so is a promising model for explaining the 
discrepancies between the measurements of $R_{K^{(\ast)}}$ and their SM predictions.  

\subsection{Example case \label{sec:eg}}
In order to identify the couplings of the model further, we need 
to specify the mixing matrices $V_I$. However, at this coarse level of model
building, we do not have an explicit model for them. 
We now make a number of (fairly strong) assumptions in order to specify a
model, but we emphasise that these just provide an example case of the model
for further study.

We know that we require a coupling
of the $Z^\prime$ to $\mu^+\mu^-$ and to $\bar b s$, in order to produce the
effective operators in Eq.~\ref{effOp}. The existence of these couplings
implies that $V_{d_L}$ and $V_{e_L}$ should contain some mixing between the
third and second generations. For now, we will
take the limiting case that 
\begin{equation}
V_{d_L}=\left( \begin{array}{ccc}
1 & 0 & 0 \\
0 & \cos \theta_{sb} & -\sin \theta_{sb} \\
0 & \sin \theta_{sb} & \cos \theta_{sb} \\
\end{array}\right)
\qquad\text{and}\qquad
V_{e_L}=\left( \begin{array}{ccc}
1 & 0 & 0 \\
0 & 0 & 1 \\
0 & 1 & 0 \\
\end{array}\right),
\label{vdl}
\end{equation}
where we expect $|\sin \theta_{sb}| \sim {\mathcal O}(|V_{ts}|)$.
Sometimes, we shall exemplify with $\sin \theta_{sb}=|V_{ts}|=0.04$ (when we
shall 
explicitly state it).
Eq.~\ref{vdl} implies that there are no tree-level flavour changing currents
between the first two generations of down quark, circumventing strong
$K^0-\bar{K^0}$ mixing constraints.\footnote{Promoting the zeroes in $V_{d_L}$
to CKM-suppressed elements does provide constraints but does not rule all of
the otherwise viable parameter space of the model out.} 
From Eq.~\ref{mix}, we require
$V_{u_L}= V_{d_L}V^\dagger$. So as not to produce $Z^\prime$ couplings to 
$\overline{b_R}s_R+H.c.$ (such couplings dominating is disfavoured by fits
to $B-$data~\cite{DAmico:2017mtc}), we set 
$V_{d_R}=1$. For simplicity and 
definiteness, we also
set $V_{u_R}=1$.
We have chosen $V_{e_L}$ in Eq.~\ref{vdl} to transfer the $Z^\prime$ coupling
from the 
third family entirely into the second in the (left-handed) charged leptons, so as to induce the
$\overline{\mu_L}\mu_L$ coupling to the $Z^\prime$. This is really a constraint upon the charged lepton
Yukawa matrix, which, up to small corrections, should then be 
\begin{equation}
Y_E = \left( \begin{array}{ccc}0 & 0 & 0 \\ 0 & 0 & Y_\tau \\ 0 & x & 0
  \\  \end{array} \right), \label{Ye}
\end{equation}
where $x$ is a Yukawa coupling contributing to the muon Dirac mass after
electroweak symmetry breaking. 
In other words, to realise this example case the 33 element of $Y_E$ must be suppressed relative to the na\"ive expectation, which presents a requirement on more detailed model building.
Note that since $Y_\tau \sim {\mathcal O}(10^{-2})$
anyway, the non-zero Yukawa couplings in Eq.~\ref{Ye} can still plausibly
result from non-renormalisable operators as required from our charge
assignment in Table~\ref{tab:charges}.\footnote{Indeed, within a
  Froggatt-Nielsen setup with flavon charge $F_\theta=\pm 1/2$, one would
  expect the coupling $Y_\tau$ to be hierarchically larger than $x$, thereby
  being consistent with the charged lepton mass hierarchy (provided, of
  course, that one can suppress the 33 element of $Y_E$ with more detailed
  model building).} In this particular form of the example case therefore, the
Third Family Hypercharge model {\em per se}\/ only explains the hierarchical
heaviness of the third family of quark: more detailed model building would be
needed to understand that of the leptons. 

Eq.~\ref{vdl} should be understood as a straightforward limiting case which fits the data at present (as we discuss in detail in \S~\ref{pheno of ex case}): 
it allows for a large coupling of the
$Z^\prime$ to muons, but kills $Z^\prime$ couplings to left-handed electrons 
which have strong constraints from LEP\@. Furthermore, with this choice there
is no $Z^\prime$ coupling to left-handed $\mu^\pm \tau^\mp$ pairs, which means this
example case is automatically consistent with very strong constraints from the
measurement of the $\tau\rightarrow \mu\mu\mu$ branching ratio
\cite{PhysRevD.98.030001}.  
Simplicity also motivates us to set 
$V_{e_R}=1$, but Eq.~\ref{mix} implies that we must set $V_{\nu_L}=V_{e_L} U^\dagger$.
 
Substituting these matrices into Eq.~\ref{secSU2}, we obtain
\begin{eqnarray}
\mathcal{L}_{X \psi} =-&g_F&\left( 
\frac{1}{6}\overline{\bf u_L} \Lambda^{(u_L)} \gamma^\rho {\bf u_L} + \frac{1}{6}
\overline{\bf d_L} \Lambda^{(d_L)} \gamma^\rho {\bf d_L}-\frac{1}{2}
\overline{\bf n_L} \Lambda^{(n_L)} \gamma^\rho {\bf n_L} -\frac{1}{2}
\overline{\mu_L} \gamma^\rho \mu_L\right. \nonumber \\
&&\left. +\frac{2}{3}
\overline{t_R} \gamma^\rho {t_R}-\frac{1}{3}
\overline{b_R} \gamma^\rho {b_R}-
\overline{\tau_R}  \gamma^\rho {\tau_R}
\right)Z^\prime_{\rho},\label{ex couplings}
\end{eqnarray}
where 
$\Lambda^{(u_L)}=V V_{d_L}^\dagger \xi V_{d_L} V^\dagger$,
$\Lambda^{(n_L)}=U V_{e_L}^\dagger \xi V_{e_L} U^\dagger$, and
\begin{equation}
\Lambda^{(d_L)} = \left( \begin{array}{ccc}
0 & 0 & 0 \\
0 & \sin^2 \theta_{sb} & \frac{1}{2}\sin 2\theta_{sb} \\
0 & \frac{1}{2}\sin 2\theta_{sb} & \cos^2 \theta_{sb} \\
\end{array} \right).
\end{equation}
From these,\footnote{Some aspects
of $Z^\prime$ couplings of the example case are somewhat similar to an ansatz
proposed in Ref.~\cite{He:2009ie}, which examined phenomenological bounds on
 them.} we 
read off the couplings relevant for causing the recent neutral 
 current lepton flavour
non-universality measurements in $B-$decays, 
\begin{equation}
\mathcal{L}_{X \psi} =-\left( 
\frac{g_F}{12} \sin 2 \theta_{sb} \overline{s} \gamma^\rho P_L{b}
-\frac{g_F}{2} \bar{\mu} \gamma^\rho P_L \mu + H.c. 
\right)Z^\prime_\rho + \ldots \label{hurrah}
\end{equation}
Eq.~\ref{hurrah} is a promising operator for explaining $R_{K^{(\ast)}}$: it
only has left-handed currents between ${\overline b} s$ and $\overline{\mu} \mu$. 
Also, the $Z^\prime$ coupling to $\bar s b$ is
suppressed by $(\sin 2 \theta_{sb})/6$
compared to its coupling to muons. This helps explain
why the model does not induce a large new physics contribution to
$B_s-\overline{B_s}$ mixing that would 
be incompatible with measurements, but can still explain $R_{K^{(\ast)}}$,
which we show in 
the next section where we examine the phenomenology of the example case. 

As one can see from Eq.~\ref{ex couplings}, the $Z^\prime$ also has couplings to all
flavours of left-handed (LH) up-type quarks, and all flavours of LH neutrinos. These couplings result in various additional constraints on the model (and predictions of the model), both through direct couplings to the $Z^\prime$ boson and via modified $Z$ couplings due to the $Z-Z^\prime$ mixing. Explicitly, $\Lambda^{(u_L)}$ has matrix elements given by
\begin{equation}
\Lambda^{(u_L)}_{ij}=\cos^2\theta_{sb}V_{ib}V_{jb}^\ast + \sin^2\theta_{sb}V_{is}V_{js}^\ast + \frac{1}{2}\sin 2\theta_{sb} (V_{is}V_{jb}^\ast + V_{ib}V_{js}^\ast),
\end{equation}
where the indices $i$ and $j$ here run over the up-type flavours $u$, $c$, and
$t$. Numerically, for $(\sin 2 \theta_{bs})/2=0.04$, the magnitudes of these
couplings are $g_F/6$ multiplied by
\begin{equation}
\Lambda^{(u_L)} = \left( \begin{array}{ccc}
0.0002 & 0.001 & 0.012 \\
0.001 & 0.006 & 0.079 \\
0.012 & 0.079 & 0.995 \\
\end{array} \right).\label{up type ex couplings}
\end{equation}

\section{Phenomenology of Example Case} \label{pheno of ex case}

We now examine the phenomenology of our Third Family Hypercharge Model example
case,
beginning with constraints, then providing predictions in terms of $Z^\prime$
width and branching ratios, and predictions for $B$ meson decays. 

\subsection{Constraints \label{sec:constraints}}
We expect the strongest constraints upon our model to come from fitting
$R_{K^{(\ast)}}$, $B_s-\overline{B_s}$ mixing measurements, and from the
measurements of $Z$ boson couplings to the first two generations of leptons
derived from LEP\@. There may be additional constraints coming from
other 
electroweak measurements: these may even provide an opportunity for our model
to better fit the forward-backward asymmetry of $b$-quarks measured by LEP,
which differs to the SM fit by some $\sim 2.3 \sigma$~\cite{PhysRevD.98.030001}. Such a study
would require a combined fit to electroweak data and is
outside the scope of the present paper. We leave it for future work, focusing now
on the other constraints in turn.

\subsubsection{Neutral current $B$ meson measurements}
A fit~\cite{DAmico:2017mtc} to 
$R_{K^{(\ast)}}$ and selected other `clean' (i.e.\ observables with
particularly low
theoretical uncertainties) $B-$observables 
found
that the couplings and mass of $Z^\prime$ particles are constrained to be 
\begin{equation}
g_{sb} g_{\mu\mu}  = -x \left(
  \frac{M_{Z^\prime}}{\text{31~TeV}}\right)^2, \label{anom}
\end{equation}
where $x=1.00 \pm 0.25$ from the fit and the relevant $Z^\prime$ couplings are
defined to be\footnote{We note that we predict a tree-level $Z$ boson
  contribution to the $(\overline{b_L} 
s_L) (\overline{\mu_L} \mu_L)$ operator in this model but since, to leading
order in 
($M_Z^2/{M_Z^{\prime}}^2$), there is an identical $Z$ contribution to $(\overline{b_L}
s_L) (\overline{e_L} e_L)$, it cancels in $R_K$ and $R_{K^\ast}$. 
}
\begin{equation}
\mathcal{L}_{Z^\prime f} =-\left( g_{sb} Z^\prime_{\rho} \overline{s_L} \gamma^{\rho} b_L  +
H.c. \right)  - g_{\mu\mu} Z_{\rho}' \overline{\mu_L}\gamma^\rho \mu_L + 
\ldots \label{coup}
 \end{equation}
From Eq.~\ref{hurrah} we identify $g_{sb}=g_F (\sin 2 \theta_{sb})/12$ and $g_{\mu\mu}=-g_F/2$ in our example case.
We can then match Eq.~\ref{anom} on to a constraint on
$g_F$ and $M_{Z^\prime}$:
\begin{equation}
g_F^2 = x \frac{24}{\sin 2 \theta_{sb}}  \left(
  \frac{M_{Z^\prime}}{\text{31~TeV}}\right)^2 =
x \left(\frac{M_{Z^\prime}}{\text{1.79~TeV}}\right)^2 \frac{0.04}{\frac{1}{2} \sin 2 \theta_{sb}}. \label{constraint}
\end{equation}
This translates to the bounds
\begin{equation}
\frac{M_{Z^\prime}}{\text{2.53~TeV}} \sqrt{\frac{0.04}{\frac{1}{2}\sin 2
    \theta_{sb}}}< g_F < \frac{M_{Z^\prime}}{\text{1.46~TeV}} 
\sqrt{\frac{0.04}{\frac{1}{2}\sin 2 \theta_{sb}}}
\end{equation}
at the 95\% Confidence Level (CL).

\subsubsection{Neutral meson mixing}
The $Z^\prime$ coupling to $b\bar{s}$ which is needed to fit $R_{K^{(\ast)}}$ also results in a tree level contribution to $B_s-\overline{B_s}$ mixing (which, in the SM, arises from box diagrams and so is loop-suppressed).
We adapt the bound on $B_s-\overline{B_s}$ mixing from
Ref.~\cite{Bazavov:2016nty,DAmico:2017mtc}, using the 
2$\sigma$ 2016 
FLAG average on the hadronic form factor $f_{B_s}$ and the bag parameter
$B_{B_s}$. 
The resulting bound is equivalent to\footnote{A recent determination of
  $f_{B_s}$ and 
  $B_{B_s}$ by the Fermilab/MILC
collaboration~\cite{Bazavov:2016nty} is in tension with other previous
estimates. When used to extract the 
Standard Model prediction of the $B_s - \overline{ B_s}$ mixing parameter $\Delta
m_s$, it is also in tension with the experimental determination. However, were
 we
to use these determinations, stronger bounds on new physics would
follow~\cite{DiLuzio:2017fdq}, implying
$|g_{sb}| \lesssim M_{Z^\prime}/600$ TeV.}
\begin{equation}
\frac{g_F}{12} \sin2\theta_{sb} <
\frac{M_{Z^\prime}}{\text{148~TeV}}
\Rightarrow 
g_F < \left(\frac{M_{Z^\prime}}{\text{1.0~TeV}} \right)
\left(\frac{0.04}{\frac{1}{2}\sin 2 \theta_{sb}}\right). \label{bsbsbar}
\end{equation}
In addition to the $Z^\prime$ contribution, there is also a tree level
contribution to $B_s-\overline{B_s}$ mixing from $Z$ boson exchange in our
model, due to the $Z-Z^\prime$ mixing. However, this contribution is
suppressed with respect to 
the $Z^\prime$ contribution by ${\mathcal O}(M_Z/M_Z^\prime)^2$ and so we
neglect it. 
Flavour-changing couplings of the $Z^\prime$ to the down-type quarks (induced by
promoting some of the zeroes in
Eq.~\ref{ex couplings} to finite quantities) would also induce corrections
beyond the SM
to the mixings of other neutral mesons, 
for example 
$K^0-\overline{K^0}$ mixing or $B_d-\overline{B_d}$
mixing~\cite{Charles:2013aka}. These would induce additional constraints on
the model. 

\subsubsection{Lepton flavour universality of the $Z$ boson}\label{LFU}
In the SM, the $Z$ boson is a linear combination of $B$ and $W^3$, {\em viz.} $Z_\mu^{\text{SM}}=\cos\theta_w W_\mu^3-\sin\theta_w B_\mu$,
whereas in the Third Family Hypercharge Model the $Z$ contains a small
admixture of the $U(1)_F$ gauge boson $X$, as in Eq.~\ref{zBoson}. Since
the fermion couplings to $X$ are flavour-dependent, this introduces
non-universality to the leptonic decays of the $Z$, which are constrained by
the LEP measurement~\cite{PhysRevD.98.030001}:
\begin{equation}
R_{\text{LEP}} =0.999\pm 0.003,
\qquad R\equiv\frac{\Gamma(Z\rightarrow e^+e^-)}{\Gamma(Z\rightarrow \mu^+\mu^-)}. \label{LEP}
\end{equation}
In the Third Family Hypercharge Model, the partial width for $Z\rightarrow e^+e^-$ is unchanged from the SM, to leading order in $\alpha_z$, because the $Z^\prime$ does not couple to (left-handed or right-handed) electrons.\footnote{There is of course a reduction in the $Z$ boson couplings to electrons arising from the factor of $\cos\alpha_z$ in Eq.~\ref{zBoson}, however this shift is of order $\alpha_z^2$ and is therefore subleading.} In contrast, the partial width for $Z\rightarrow \mu^+\mu^-$ {\em is}\/ modified at leading order, because of the $X$ coupling to left-handed muon pairs.

Within the Third Family Hypercharge Model, the ratio of partial widths is
\begin{equation}
R_{\text{model}} = \frac{|g_Z^{e_L e_L}|^2+ |g_Z^{e_R e_R}|^2}{|g_Z^{\mu_L \mu_L}|^2+|g_Z^{\mu_R \mu_R}|^2},
\end{equation}
where $g_Z^{ff}$ is the coupling of the physical $Z$ boson to the fermion pair
$f\bar f$. One can obtain the couplings $g_Z^{ff}$ by first writing down the terms
in the Lagrangian which couple the charged leptons to the neutral bosons $B$,
$W^3$, and $X$:
\begin{eqnarray}\mathcal{L}_{l Z^\prime}&=&-\overline{e_L}\left(-\frac{1}{2}g
\slashed{W}^3-\frac{1}{2}g'\slashed{B}\right) e_L -
\overline{\mu_L}\left(-\frac{1}{2}g \slashed{W}^3-\frac{1}{2}g'\slashed{B}-\frac{1}{2}g_F\slashed{X}\right) \mu_L - \nonumber \\ &&
\overline{\tau_L}\left(-\frac{1}{2}g\slashed{W}^3-\frac{1}{2}g'\slashed{B}\right)\tau_L -\overline{{\bf                                                                                                                                  e_R}}\left(-g'\slashed{B}\right){\bf
                                            e_R}- \overline{\tau_R} \left(
                                                                                                     -g_F \slashed{X}                                     \right) \tau_R
,
\end{eqnarray}
and then inserting ${\bf A_\mu}'=O {\bf A_\mu}$ (where $O$ is given in
Eq.~\ref{orthogonal}) to rotate to the mass basis. To leading order in 
$\sin\alpha_z$), we find: 
\begin{eqnarray}
\begin{aligned}
g_Z^{e_L e_L} &=-\frac{1}{2}g\cos\theta_w+\frac{1}{2}g'\sin\theta_w,\\
g_Z^{\mu_L \mu_L} &=-\frac{1}{2}g\cos\theta_w+\frac{1}{2}g'\sin\theta_w-\frac{1}{2}g_F\sin\alpha_z,\\
g_Z^{e_R e_R} &= g_Z^{\mu_R \mu_R}=g'\sin\theta_w.
\end{aligned}
\end{eqnarray}
The SM prediction ({\em i.e.} $R=1$) is
recovered by taking $\alpha_z$ to zero. Within the Third Family
Hypercharge Model, we may expand $R_{\text{model}}$ to leading order in
$\sin\alpha_z$: 
\begin{equation}
R_{\text{model}} = 1 -\frac{2 g_F(g\cos\theta_w-g'\sin\theta_w)\sin\alpha_z}{(g\cos\theta_w-g'\sin\theta_w)^2+4g'^2\sin^2\theta_w}
= 1-4.2 g_F^2 \left(\frac{M_Z}{M_{Z^\prime}}\right)^2, 
\end{equation}
after substituting in Eq.~\ref{mixing} for $\sin\alpha_z$, and the central experimental values $g=0.64$ and $g'=0.34$.
Comparison with the lower LEP limit, at the $95\%$ CL,
yields the $Z-$ LEP lepton 
flavour universality constraint (LEP LFU) from Eq.~\ref{LEP}:
\begin{equation}
g_F^2 \left(\frac{M_Z}{M_{Z^\prime}}\right)^2< 0.0017 \Rightarrow
g_F < \frac{M_{Z^\prime}}{\text{2.2~TeV}}. \label{LEPLFU}
\end{equation}
Other constraints from LEP measurements of fermionic couplings to $Z$ bosons (for
example $b \bar b$) are weaker than this. We see that LEP LFU yields a 
tighter constraint
than the one from $B_s-\overline{ B_s}$ in Eq.~\ref{bsbsbar} for
$\frac{1}{2}\sin 2\theta_{sb} \lesssim 0.08$.

\subsubsection{$t \rightarrow Z q$ Decays}

One might worry that in our example case we have introduced various tree level
flavour changing neutral current interactions involving the top quark and the
 lighter up-type quarks $u$ 
and $c$. However, it turns out that constraints on our model from
flavour-changing $tZ$ interactions are very weak, as we now summarise for
completeness. In the example case of the Third Family Hypercharge Model, the
Lagrangian in Eq.~\ref{ex couplings} contains the interactions\footnote{As can
  be seen from Eq.~\ref{up type ex couplings}, the interactions involving only
  $u$ and $c$ are extremely suppressed, because the $X$ couplings are ``mixed
  in'' from the third family in the Third Family Hypercharge Model.}  
\begin{equation}
\mathcal{L}_{Xtq} =
-\frac{g_F}{6}\left(\Lambda^{(u_L)}_{23} \bar{c}\gamma^\rho P_L t + \Lambda^{(u_L)}_{13} \bar{u}\gamma^\rho P_L t  +  H.c.\right) X_\rho,
\end{equation}
(where $\Lambda^{(u_L)}_{23}\approx V_{cb}V_{tb}^\ast+\frac{1}{2}\sin
2\theta_{sb}V_{cs}V_{tb}^\ast$ and $\Lambda^{(u_L)}_{13}\approx
V_{ub}V_{tb}^\ast+\frac{1}{2}\sin 2\theta_{sb}V_{us}V_{tb}^\ast$), 
facilitating the decays $t\rightarrow Zu$ and $t\rightarrow Zc$ at tree-level
via the $Z-Z^\prime$ mixing. 
In the example case, the branching ratio for $t\rightarrow Zc$ is predicted to be
\begin{eqnarray}
BR(t\rightarrow Zc)&=&\frac{ g_F^2 \Lambda^{(u_L)2}_{23}f(M_Z,M_W,M_t) \sin^2\alpha_z}{18 g^2|V_{tb}|^2}BR(t\rightarrow Wb) \nonumber \\ 
&&=1.1 \times 10^{-3}g_F^4 \left( \frac{M_Z}{M_{Z^\prime}} \right)
   ^4\left(\frac{|V_{cb}V_{tb}^\ast+\frac{1}{2}\sin
   2\theta_{sb}V_{cs}V_{tb}^\ast|^2}{0.0062}   \right), \label{tDK}
\end{eqnarray}
where $f(M_Z,M_W,M_t)$ is a kinematical
factor,\footnote{Explicitly, $$f(M_Z,M_W,M_t)=\frac{M_W^2}{M_Z^2}\left(1-\frac{M_Z^2}{M_t^2}\right)^2\left(1+\frac{2M_Z^2}{M_t^2}\right)\left(1-\frac{M_W^2}{M_t^2}\right)^{-2}\left(1+\frac{2M_W^2}{M_t^2}\right)^{-1},$$
  where we have neglected the masses of the bottom and charm quarks. We have
  used $M_W=80.4$ GeV, $M_Z=91.19$ GeV and $M_t=173$ GeV in evaluating the
  right-hand side of Eq.~\ref{tDK}. 
} and we have assumed the top's branching ratio to $Wb$ is unity and neglected 
the masses of the bottom and charm quarks.
Using the CMS bound from the 8 TeV LHC data, $BR(t\rightarrow Zc)< 4.9\times
 10^{-4}$ at 95\% CL \cite{Sirunyan:2017kkr}, yields the weak constraint
 $g_F< M_{Z^\prime}/(0.1\text{~TeV})$ when $\frac{1}{2}\sin
 2\theta_{sb}=0.04$. Performing a similar 
 calculation for $t\rightarrow Zu$ using the CMS 8 TeV 95\% CL bound,
 $BR(t\rightarrow Zu)< 2.2\times 10^{-4}$, yields a yet weaker constraint on
 $g_F/M_{Z^\prime}$. 

\subsubsection{Combination of constraints}
\begin{figure}
\begin{center}
\unitlength=15cm
\begin{picture}(1,0.45)(0,0)
    \put(-0.25,-0.05){\includegraphics[width=0.9\textwidth]{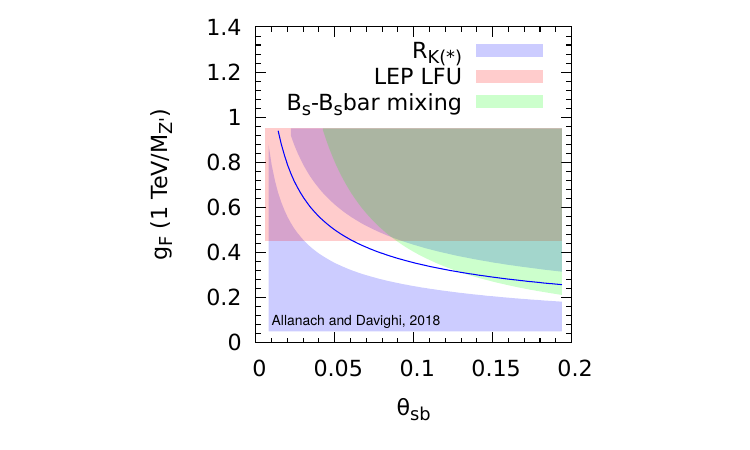}}
    \put(0.3,-0.05){\includegraphics[width=0.9\textwidth]{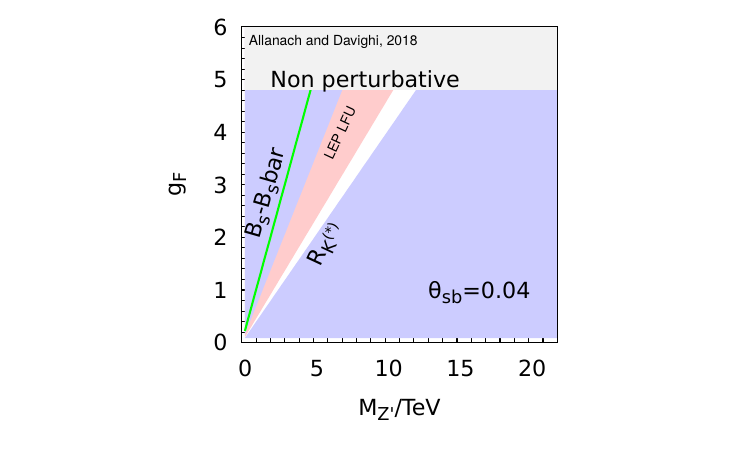}}
\end{picture}
\end{center}
\caption{Summary of 95$\%$ CL constraints upon the Third Family Hypercharge Model
  example case. Each constraint excludes the labelled coloured area, leaving
  an allowed white region. $R_{K^{(\ast)}}$ refers to a fit to `clean'
  $B-$physics observables~\cite{DAmico:2017mtc}, including to $R_K$ and
  $R_{K^{(\ast)}}$. The blue curve in the LH plot is where the fit is central.
  `LEP LFU' shows the LEP lepton flavour universality
  constraint in Eq.~\ref{LEPLFU}, whereas 
    `$B_s-B_s\text{bar}$' shows the constraint in
    Eq.~\ref{bsbsbar}. In the right-hand plot, we show constraints in the
    $g_F-M_{Z^\prime}$ plane for the choice $\theta_{sb}=0.04$. In this plot, the
    region ruled out by the $B_s-\overline{B_s}$ mixing constraint is {\em
      above}\/ the line. 
  The `Non perturbative' region is defined
  as the region where $\Gamma_{Z^\prime}>M_{Z^\prime}$ and is estimated in
  \S~\ref{sec:pred}.\label{fig:sum}}
\end{figure}

We summarise the constraints on our example case in Fig.~\ref{fig:sum}.
We see in the LH panel the white  region of parameter space, which fits
$R_{K^{(\ast)}}$ 
  whilst remaining on the right side of the LEP LFU
  and 
  $B_s-\overline{ B_s}$ mixing constraints. 
It is encouraging that our  example case can satisfy
all bounds when the fermion mixings are tightly constrained 
by rather simple and definite choices. The central value of the fit to clean
$B-$physics observables can be achieved for $0.13 \geq \theta_{sb}\geq 0.06$,
as shown by 
the blue line in the LH panel, and the constraints.
If we were to choose more general fermion
mixing matrices, 
we might widen the allowed region. For now, we leave the example case as an
existence proof. 

\subsection{Predictions \label{sec:pred}}

Here, we sketch the main experimental predictions of the model.
The $Z^\prime$ particle may be able to be produced and
measured~\cite{Allanach:2017bta} 
either at the LHC, the
high luminosity run of the LHC, or the high energy run of the LHC or a 100 TeV
future circular $pp$ collider
in order to
provide a direct test of the Third Family 
Hypercharge Model, and hence of the mechanism that generates the
hierarchically heavy third family of charged fermions. 
The classic signature of such a $Z^\prime$ would be a bump in the muon
anti-muon pair production cross-section~\cite{Allanach:2017bta}.
Current searches by LHC general purpose experiments have so far found no such
bump, for example excluding $M_{Z^\prime}<4.0,4.5$ TeV for $Z^\prime$ models 
that couple 
identically to the SM $Z$ boson~\cite{Sirunyan:2018exx,Aaboud:2017buh}. The
SM $Z$ boson has sizeable couplings to up and down quarks, whereas in our example
case the $Z^\prime$ has only very small quark couplings, except for the third
generation: the bounds from 
such direct searches are then very much less sensitive than the 4.0-4.5 TeV
masses quoted. We
leave the re-casting of LHC bounds for our model to future work. 

\begin{table}
\begin{center}
\begin{tabular}{|cc|cc|cc|}
\hline 
Mode & BR & Mode & BR & Mode & BR \\ \hline
$t\bar t$ & 0.42 & $b \bar b$ & 0.12 & $\nu \bar \nu'$ & 0.08 \\
$\mu^+ \mu^-$ & 0.08 & $\tau^+ \tau^-$ & 0.30 & other $f_i f_j$ & $\sim
                                                                  {\mathcal O}(10^{-4})$ \\
\hline
\end{tabular}
\end{center}
\caption{\label{tab:BRs} $Z^\prime$ branching ratios (BRs) in the Third Family
 Hypercharge 
  Model example case. We have neglected fermion masses and lumped all flavours
  of neutrino into $\nu$,   $\nu'$. The BRs to other fermion pairs are highly
  suppressed; for example, the next largest BR is to $t\bar{c}$ pairs, for
  which the BR is $\sim {\mathcal O} (10^{-4})$.
  }  
\end{table}
Assuming that $M_{Z^\prime} \gg 2 m_t$ (since
otherwise it would likely have been discovered already), we may neglect  
fermion masses in its decays. The partial width of $Z^\prime$ into 
a massless
fermion $f_i$
and anti-fermion $\bar f_j$ is $\Gamma_{f_i f_j}=C/(24 \pi) |g_{ij}|^2
M_{Z^\prime}$, where $g_{ij}$ is the coupling of the $Z^\prime$ to $f_i
\bar f_j$, and 
$C=3$ if the fermions are coloured but $C=1$ otherwise. $Z^\prime$ has a
tree-level coupling to $h Z$ in our model, which stems from giving the Higgs a $U(1)_F$
  charge. Specifically, upon rotation
  to the mass basis of the neutral gauge bosons, as in \S~\ref{ZZpmixing}, one finds the Lagrangian
  terms in Eq.~\ref{scalarKineticTerms} contain a piece
  $-g_F M_Z \lambda hZ^{\prime}_\mu Z^\mu$ where we
  calculate the coefficient to be
  $\lambda = -g_F M_Z ( 1 + \mathcal{O}(M_Z/M_{Z^\prime})^2)$.
  Thus, the partial width
  $Z^{\prime}$ into $hZ$ is
  $\propto M_{Z^\prime} / (8 \pi) \mathcal{O}(M_Z/M_{Z^\prime})^2$ and so is 
  negligible compared to decays into fermions.
Neglecting this mode,  neglecting fermion masses and working from the weak
eigenbasis 
couplings in Eq.~\ref{Zpcouplings}, we obtain a total Third Family Hypercharge
Model $Z^\prime$ 
width of 
\begin{equation}
\Gamma_{Z^\prime}=\frac{5 g_F^2 M_{Z^\prime}}{36 \pi},\label{width}
\end{equation}
where the branching ratio into quarks is 11/20 and the branching ratio
into leptons is 9/20. The $Z^\prime$ total
width is equal to its mass when $g_F=6 \sqrt{\pi/5}=4.8$. For
$g_F$ values of this size and above, the
model enters a non-perturbative r\'{e}gime, which is indicated in Fig.~\ref{fig:sum}.
In order to further specify the $Z^\prime$ branching ratios into different
flavours of 
quark and lepton, one must specify the $V_I$ mixing matrices. 

For the example case that we have detailed in \S\ref{sec:eg}, the branching ratios are as
in Table~\ref{tab:BRs}, where we have taken the central values of CKM and
PMNS matrix elements from the Particle Data Group~\cite{PhysRevD.98.030001}.
We see from Table~\ref{tab:BRs} that the example case predicts that a bump in
the 
$\mu^+\mu^-$ invariant mass spectrum at $M_{Z^\prime}$ is suppressed by the
$\approx 8\%$ branching ratio. Other promising discovery modes are likely
to then be into boosted top pairs and tau pairs (for which the branching
ratios are bigger because of the larger couplings of the $Z^\prime$ to
{\em right}-handed taus). It will be interesting to 
compare sensitivities to the different channels quantitatively, in the
future, and to estimate the sensitivities to measuring the top and tau
polarisations, which are different to those produced by $Z$ bosons.

The example case predicts a  non-SM contribution to 
${BR(B\rightarrow K^{(\ast)} \tau^+ \tau^-)}$. As such, it follows some of the
expectations from Ref.~\cite{Glashow:2014iga}.
Identifying $\tau$ particles resulting from $B$ decays is difficult
experimentally, and 
so we may have to wait for future LHC and Belle II runs before there is
sufficient sensitivity to these modes. Near future prospects for improving and
checking the 
measurements of $R_K$ and
$R_K^{(\ast)}$ remain very good, however~\cite{Albrecht:2017odf,Kou:2018nap}.

We have left the study of the Higgs potential including the flavon $\theta$
for the future. However, a gauge invariant term in the potential
$\lambda_{\theta H}
|\theta|^2 |H^2|$ is present at the renormalisable
level, where $\lambda_{\theta H}$ is a dimensionless coupling. Since $\theta$
 and $H$ both acquire VEVs, this term will induce 
flavon-Higgs mixing. 
This could then affect Higgs couplings, particularly to taus, tops and bottom
quarks. It is clear that one can remove these effects in the limit 
$\lambda_{\theta H} \rightarrow 0$, but applying current experimental bounds
on Higgs branching ratios 
would provide an upper limit on $\lambda_{\theta H}$.

In the present paper, we have focused on the tree-level phenomenology. There
are small effects at the one-loop level, for example due to 
$U(1)_Y - U(1)_F$ mixing (and indeed from Higgs-flavon mixing, even in the
$\lambda_{\theta H} \rightarrow 0$ limit), that are beyond the scope of
our paper but may be interesting to address nonetheless.

\section{Conclusions}
We have constructed a model with a gauged flavoured $U(1)_F$ group. Once it
 has been spontaneously broken via  
$$SU(3)\times SU(2)_L \times U(1)_Y
\times U(1)_F 
\stackrel{\mathclap{\theta}}{\rightarrow}
SU(3)\times SU(2)_L \times U(1)_Y \stackrel{\mathclap{H}}{\rightarrow} SU(3)
\times U(1)_{em},$$
our model explains some coarse features of fermion masses and mixings and it
provides an explanation for inferred non-SM contributions to $R_{K^{(\ast)}}$. 
In particular, the hierarchical heaviness of the third family of charged
fermions is predicted. We imagine that small non-renormalisable operators will
induce quark mixing and masses for the lighter two families. 
CKM mixing will then be predicted to be small.
PMNS mixing, however, is not necessarily predicted to be small. For example, 
in our example case we induce a large 23 PMNS mixing by requiring the 33
element of the charged lepton Yukawa be suppressed. More generally, in any
implementation of the Third Family Hypercharge Model, large PMNS mixing can
result from the neutrino sector, which we shall now discuss. 

The only dimension 5 term
allowed by the $SU(3)\times SU(2)_L \times U(1)_Y\times
U(1)_F$ gauge symmetry in the fermion sector is
$
{\mathcal L}_{SS}=\frac{1}{2M} ({L_3'}^T H^c) (L_3' H^c),
$
leading to a third family neutrino mass after $H$ develops a VEV\@. The first
two neutrino masses are not present at this order in the effective field theory expansion, being banned by $U(1)_F$, and so one
might {\em prima facie}\/ expect the 
model to predict a normal hierarchy and small mixing in the neutrino sector.  
However, once a more detailed model for non-renormalisable terms is built (for
example by including right-handed neutrinos), we
may expect the effective dimension 5 terms to instead be $
1/2 \sum_{ij} (L_i' H^c) (M^{-1})_{ij} (L_j' H^c),
$ where now $(M^{-1})_{ij}$ may well have a non-trivial structure depending on
details of the model. If some of the elements of $(M^{-1})_{ij}$ may be predicted to
be of the same order of magnitude, then an explanation for large PMNS mixing
can result. 
By extending the model with right-handed neutrinos in such a way, and by a careful
    assignment of charges and implementation of the Froggatt-Nielsen
    mechanism, we would like to provide a full account of {\em both}\/ the mass hierarchies
    and the mixing angles in all the SM fermions. We intend to explore such 
    avenues in the future. 

We emphasise here that the example case presented in \S~\ref{sec:eg} contains some
very strong assumptions, where various limits are taken for definiteness. 
One would ideally want to derive the structure from the Froggatt-Nielsen (or
similar) mechanism, and thereby to develop a more refined example case for
study. However, the example case which we have set up in this paper helps eke out phenomenological predictions in
a particular limit. Suggestions for future measurements issuing from this include
bounding $B$ decays to $K^{(\ast)}\tau^+\tau^-$, searching for the $Z^\prime$
in boosted top pairs,  di-taus and 
 di-muons, and high luminosity LHC searches for {\em e.g.} $t\rightarrow Zc$
 decays. 

One can imagine variants of the model. One such variant would be to
make the $Z^\prime$ only couple to the $\mu$ flavour of charged lepton, by switching
the second family lepton hypercharges with those of the third family:
$F_{L'_3}=F_{{e_R}'_3}=0$, $F_{L'_2}=-1/2$, $F_{{e_R}'_2}=-1$, with all other
charges as in Table~\ref{tab:charges}.\footnote{Such a $U(1)_F$ charge
  assignment remains anomaly-free.} In this case, the tau Yukawa coupling 
would be absent from Eq.~\ref{3famYuk}, and so it would need to be produced by
an non-renormalisable operator with effective coefficient $\approx
{\mathcal O}(10^{-2})$. On the other hand, the 
muon Yukawa coupling {\em would}\/ be 
present at the renormalisable level and would need to be set to be fairly
small: $\mathcal{O}(m_\mu / 
m_t)\sim 10^{-3}$. In this case, one could fix $V_{e_L}=1$ meaning that
{\em all}\/ of the PMNS mixing would come from the neutrinos,
 $V_{\nu_L}=U^\dagger$. The LEP LFU constraint in Fig.~\ref{fig:sum} would no
 longer apply, widening the parameter space shown in the figure. 
The $Z^\prime$ would 
then couple only to quarks, neutrinos and muons (i.e.\ not to
$\tau \tau$). This tweaked model is similar to the `$33\mu\mu$'  
model of Ref.~\cite{Allanach:2017bta}, except that the $Z^\prime$ would
contain additional couplings to $\mu_R$ as well as to $\mu_L$. 

There are discrepancies with SM predictions at a similar level to
$R_{K^{(\ast)}}$ (when
measured in numbers of sigma) in
$B\rightarrow D^{(\ast)}\tau \nu$ decays~\cite{Lees:2013uzd,Aaij:2015yra,Hirose:2016wfn,Aaij:2017deq}, which we have not addressed in our
model. However, to explain this charged 
current, a different mass scale is required to the one that explains
$R_{K^{(\ast)}}$: the mass 
divided by the square root of the product of two of its couplings is required
to be around 3.4 
  TeV~\cite{DAmico:2017mtc} in order to fit the data, i.e.\ an order of
  magnitude lighter, or a more 
  strongly coupled particle. 
We note some ambitious models explaining,  to some
extent, both the $B\rightarrow D^{(\ast)}\tau
\nu$ data and $R_{K^{(\ast)}}$~\cite{Bordone:2017bld,Greljo:2018tuh} based on gauged
vector leptoquarks\footnote{We notice the appearance of a Pati-Salam vector
 leptoquark in Ref.~\cite{Assad:2017iib} in order to explain the discrepant $B-$measurements.}~\cite{DiLuzio:2017vat,Buttazzo:2017ixm}. These models are rather
involved (for example, they contain both a $Z^\prime$
and a leptoquark). 
We have limited the scope of our much simpler model and we ignore
  the $B\rightarrow D^{(\ast)}\tau \nu$ data, being 
  content for now to explain only the neutral current discrepancies with SM
  predictions. If the charged current $B$ discrepancies 
  stand the test of time, clearly the model should be extended in order to
  take them   into account.

The Third Family Hypercharge Model explains the $R_{K^{(\ast)}}$ measurements by predicting a $Z^\prime$
with flavour dependent couplings. The third family of fermions (and the Higgs
doublet) has a $U(1)_F$ charge given by the hypercharge, resulting in
a hierarchically massive third family of fermions and small CKM mixing
elements. The precise 
constraints upon the model do depend upon choices in the fermion mixing
parameters. We have showed, in one simple example case, an existence proof
where the model passes the relevant current experimental tests.
The model as a whole is fairly concise, requiring no additional
fields to the SM, save for the 
$U(1)_F$ gauge field and a complex SM singlet scalar to spontaneously break the
symmetry, and is 
moreover anomaly free. 
\begin{quote}
The Third Family Hypercharge Model (and other models of its ilk), raise the
exciting 
possibility of {\em providing a direct experimental probe
(through measurements of $Z^\prime$ couplings) of mechanisms pertinent to the
`fermion masses and mixings' problem}.
\end{quote}
\section*{Acknowledgements}
We thank other members of the Cambridge SUSY Working Group,
M AlTakach, F Coradeschi, A Celis, A
 Crivellin, T You and 
 M Nardecchia for their helpful advice 
and comments and to C Thomas and M Wingate for advice on flavour bounds.
BCA thanks the CERN TH Unit for hospitality offered while the project was in
its final stages. 
This work has been partially supported by STFC consolidated grant
ST/P000681/1. JD has been supported by The Cambridge Trust.

\appendix

\section{Fermion Rotation to the Mass Basis \label{sec:weakToMass}}
Here, we detail the rotation of SM fermion fields to the mass basis in order to
fix our conventions.
We write 
$$
{\bf u_L'}=\left( \begin{array}{c} u_L' \\ c_L' \\ t_L' \\ \end{array}
\right), \qquad
{\bf d_L'}=\left( \begin{array}{c} d_L' \\ s_L' \\ b_L' \\ \end{array} \right), \qquad
{\bf n_L'}=\left( \begin{array}{c} {\nu_e}_L' \\ {\nu_\mu}_L' \\ {\nu_\tau}_L' \\ \end{array} \right), \qquad
{\bf e_L'}=\left( \begin{array}{c} e_L' \\ \mu_L' \\ \tau_L' \\ \end{array}
\right),\qquad
$$
$$
{\bf u_R'}=\left( \begin{array}{c} u_R' \\ c_R' \\ t_R' \\ \end{array}
\right), \qquad
{\bf d_R'}=\left( \begin{array}{c} d_R' \\ s_R' \\ b_R' \\ \end{array} \right), \qquad
{\bf e_R'}=\left( \begin{array}{c} e_R' \\ \mu_R' \\ \tau_R' \\ \end{array}
\right),\qquad
$$
along with the SM fermionic electroweak doublets
$$
{\bf Q_L'}_i=\left( \begin{array}{c} {\bf u_L'}_i \\ {\bf d_L'}_i \end{array}
\right),\qquad
{\bf L_L'}_i=\left( \begin{array}{c} {\bf n_L'}_i \\ {\bf e_L'}_i \end{array}
\right).  
$$
The SM fermions acquire their mass through the terms
\begin{equation}
-\mathcal{L}_{Y}=\overline{\bf Q'_L}  Y_u H {\bf u'_R} +
\overline{\bf Q'_L}  Y_d H^c  {\bf d'_R} +
\overline{\bf L'_L}  m_e H^c  {\bf e'_R} + \frac{1}{2}
(\overline{{\bf L'_L}^c} H^\dag)  M^{-1}  ({\bf L'_L}H^\dag) + H.c., \label{yuk}
\end{equation}
where $Y_u$, $Y_d$ and $Y_e$ are dimensionless complex coupling constants,
each written as a 3 by 3 matrix in family space. These will have large 33 
elements and smaller off-diagonal elements, in agreement with
Eq.~\ref{3famYuk}. The matrix $M^{-1}$ is a 3 by 3 matrix of mass dimension
-1 and ${\bf L'_L}^c$ is the
charge conjugate of the vector of left-handed lepton doublets.
After electroweak symmetry breaking, the terms in Eq.~\ref{yuk} become the
fermion mass terms plus some Higgs interactions:
\begin{eqnarray}
-\mathcal{L}_{Y}&=&\overline{\bf u'_L} V_{u_L} V_{u_L}^\dagger m_u V_{u_R}
V_{u_R}^\dagger {\bf u'_R} +
\overline{\bf d'_L} V_{d_L} V_{d_L}^\dagger m_d  V_{d_R} 
V_{d_R}^\dagger {\bf d'_R} + \nonumber \\ &&
\overline{\bf e'_L} V_{e_L} V_{e_L}^\dagger m_e  V_{e_R} 
V_{e_R}^\dagger {\bf e'_R} + 
\overline{{\bf n'_L}^c}  V_{{\nu}_L}^* V_{{\nu}_L}^T
                                            m_\nu  V_{{\nu}_L} 
V_{{\nu}_L}^\dagger {\bf n'_L} + H.c. +\ldots
\end{eqnarray}
where $V_{I_L}$ and $V_{I_R}$ are 3 by 3 unitary matrices for each species $I$, ${\bf n'_L}^c$ is
the charge conjugate of the left-handed neutrino field, $m_u=v Y_u$, $m_d=v
Y_d$, $m_e=v Y_e$, and $m_\nu$ is the effective Majorana light neutrino mass
matrix.

Choosing
$V_{I_L}^\dagger m_I  V_{I_R}$ to be diagonal, real and positive for the $I
\in \{ u,d,e\}$, and
$V_{{\nu}_L}^T m_\nu  V_{{\nu}_L}$ to be diagonal, real and positive for the
neutrinos
(all in increasing order of mass
toward the bottom right of the matrix), we can identify the {\em non}\/ primed mass
eigenstates
\begin{eqnarray}
{\bf u_R}\equiv V_{u_R}^\dagger {\bf u_R}', \qquad &
{\bf u_L}\equiv V_{u_L}^\dagger {\bf u_L}', \qquad &
{\bf d_R}\equiv V_{d_R}^\dagger {\bf d_R}', \qquad
{\bf d_L}\equiv V_{d_L}^\dagger {\bf d_L}',  \nonumber \\
{\bf e_R}\equiv V_{e_R}^\dagger {\bf e_R}', \qquad &
{\bf e_L}\equiv V_{e_L}^\dagger {\bf e_L}', \qquad &
{\bf n_L}\equiv V_{\nu_L}^\dagger {\bf n_L}'. \label{fermion rotations}
\end{eqnarray} 
We may then identify the Cabibbo-Kobayashi-Maskawa matrix (CKM) $V$ and the
Pontecorvo-Maki-Nakagawa-Sakata (PMNS) matrix $U$:
\begin{equation}
V=V_{u_L}^\dagger V_{d_L}, \qquad U = V_{\nu_L}^\dagger V_{e_L}. \label{mix}
\end{equation}

\bibliographystyle{JHEP-2}
\bibliography{articles_bib}

\end{document}